\documentstyle[aps,prl,multicol,psfig]{revtex}
\tighten
\begin{document}
%\draft
\title{Vortex, skyrmion and elliptical domain wall textures 
 in the two-dimensional Hubbard model}
\author{G. Seibold}
\address{Institut f\"ur Physik, BTU Cottbus, PBox 101344, 
         03013 Cottbus, Germany}
\date{\today}
\maketitle

\begin{abstract}
The spin and charge texture around doped holes in the two-dimensional
Hubbard model is calculated within an unrestricted 
spin rotational invariant slave-boson approach. In the first part we 
examine in detail the spin structure around two holes doped in the
half-filled system where we have studied cluster sizes up
to $10 \times 10$. It turns out that the most stable configuration 
corresponds to a vortex-antivortex pair which has lower energy than
the N\'{e}el-type bipolaron even when one takes the far field contribution 
into account. We also obtain  
skyrmions as local minima of the energy 
functional but with higher total energy than the vortex solutions.
Additionally we have investigated the stability of
elliptical domain walls for commensurate hole concentrations. 
We find that 
(i) these phases correspond to local minima of the energy functional only 
in case of partially filled walls, (ii) elliptical domain walls are 
only stable in the low doping regime.
\end{abstract}

\vspace*{0.2cm}


\begin{multicols}{2}

\section{Introduction}
The description of inhomogeneous charge and spin phases in the
Hubbard model it a topic of current interest, mainly because
it is now generally accepted that the high-T$_c$ compounds, at least in the
underdoped regime, are intrinsic electronically inhomogeneous systems.   
A powerful tool for the investigation of such inhomogeneous electronic
states is the unrestricted Hartree-Fock (HF) scheme which allows for
the diagonalization of reasonable cluster sizes. 
Recently an extension of this approach based on the Gutzwiller wave function
\cite{GSH} has shown to significantly improve the HF solutions which
strongly underestimate the attraction between charge carriers. 
In this paper we extend the approach of Ref. \cite{GSH} to include
transversal spin degrees of freedoms which permits the description
of coplanar and three-dimensional inhomogeneous spin textures.

Among coplanar spin structures homogeneous spiral solutions have
been studied by a large variety of methods (see e.g. \cite{FLECK,FRESARD})
which show that a small amount of holes doped into the half-filled system leads
to a ${\bf Q}\sim (1,1)$ spiral phase which changes its direction
to ${\bf Q}\sim (1,0)$  above some critical concentration.
Relaxing the constraint of a homogeneous charge distribution the formation
of coplanar, vortexlike phases has been investigated in Ref. \cite{BISHOP}
using an unrestricted HF approach. These are configurations
where the antiferromagnetic (AF) spin order rotates by multiples 
of 2$\pi$ around the localized
hole. Due to this twist in the magnetization their energy increases 
$\sim ln(L^2)$ which implies their instability in large clusters.

A three-dimensional spin texture which is known to be topologically
stable is the skyrmion as a solution of the
O(3) non-linear $\sigma$-model \cite{BELAVIN}.
It has been studied by several authors also for an AF background
\cite{SHRAIMAN,GOODING} mainly concentrating on small clusters.

However, it is still controversal wether skyrmion solutions exist
also on discrete lattices. Whereas unrestricted HF theory for the 2-d Hubbard
model predicts the decay of skyrmions into conventional 
spin-polarons \cite{BISHOP}, exact diagonalization studies of a 
small cluster within the tJ-model Ref. \cite{HAAS} seem to support their
existence even when one takes the contribution of the 
skyrmion far field into acount. Within our slave-boson approach we will 
show below that skyrmions textures are stable solutions of the
discrete two-dimensional Hubbard model, when the system is doped with two
holes away from half-filling.

Connected with the discovery of static stripe order in La$_{2}$NiO$_4$ and 
La$_{1.48}$Nd$_{0.4}$Sr$_{0.12}$CuO$_{4}$ \cite{TRAN,TRAN1} a lot of work 
has been done in order to understand the different domain wall structures in  
these compounds \cite{VARIOUS,OLES,SCCG,WHITE}. 
Whereas in the Ni-doped compounds one finds
the stripes along the diagonal with one hole per Ni-site the charge
and spin order in the Nd-doped cuprates is along the copper-oxygen bond
direction with one hole per every second copper site only.
Since half-filled horizontal walls are at odds with HF calculations 
the inclusion of correlations turns out to be a necessary ingredient 
for the study of domain walls in these systems \cite{SCCG,WHITE}. 
In addition long-range Coulomb interactions
can play an important role in stabilizing half-filled vertical stripes 
\cite{SCCG}. 

Based on a Landau free-energy analysis of coupled charge and 
spin-density-wave order parameters it has been shown in Ref. \cite{ZACHAR}
that within some region of parameter domain walls may have a spiral component.
Whereas this type of ordering is not
observed in the Ni-oxides, some spiral contribution cannot be rigorously 
excluded to be present in the Nd-doped compounds \cite{TRAN1}.
According to our analysis presented below, elliptical stripes are
not stable for concentrations around $1/8$ but may be formed in
the very low doping limit.

The rest of the paper is organized as follows: In Sec. II we give a
detailed description of the formalism, in Sec. III we present
the results for vortex, skyrmion and elliptical domain wall solutions 
respectively,
and in Sec. IV we summarize our conclusions.

\section{Model and Formalism}
We consider the two-dimensional Hubbard model on a square lattice, with 
hopping restricted to nearest neighbors (indicated by the bracket $<i,j>$)
\begin{equation}\label{HM}
H=-t\sum_{<ij>,\sigma}c_{i,\sigma}^{\dagger}c_{j,\sigma} + U\sum_{i}
n_{i,\uparrow}n_{i,\downarrow}
\end{equation}
where $c_{i,\sigma}^{(\dagger)}$ destroys (creates) an electron 
with spin $\sigma$ at site
i, and $n_{i,\sigma}=c_{i,\sigma}^{\dagger}c_{i,\sigma}$. U is the
on-site Hubbard repulsion and t the transfer parameter. For the calculations
in Sec. III we take t=1.
In the following we use a spin-rotation-invariant form \cite{WH} of the 
slave-boson representation introduced by Kotliar and Ruckenstein 
in Ref. \cite{KOTLIAR}.
The subsidiary boson fields $e_{i}^{(\dagger)}$, $d_{i}^{(\dagger)}$ stand
for the annihilation (creation) of
empty and doubly occupied sites, respectively, whereas the matrix 
\begin{equation}\label{PI}
{\bf p_{i}} = \left( \begin{array}{cc} 
p_{i,\uparrow} & \frac{1}{\sqrt{2}}(p_{i,x}-i p_{i,y}) \\
\frac{1}{\sqrt{2}}(p_{i,x}+i p_{i,y}) & p_{i,\downarrow} \end{array}\right)
\end{equation}
represents the case of a singly occupied site.
Since we consider
the mean-field limit all boson operators will be approximated as numbers.
Besides the completeness condition
\begin{equation}\label{CONST1}
e_{i}^{2}+tr({\bf p_{i,\mu}^{*}p_{i,\mu}})+d_{i}^{2}=1
\end{equation}
the boson fields are constrained by the following relations
\begin{equation}\label{CONST2}
tr({\bf \tau_{\mu} p_{i}^{*}p_{i}})+2 \delta_{\mu,0} d_{i}^{2}
=\sum_{\sigma \sigma'}c_{i,\sigma}^{\dagger}({\bf \tau_{\mu}})
_{\sigma \sigma'}  c_{i,\sigma'}
\end{equation}
where ${\bf \tau_{\mu=1,2,3}}$ are the Pauli spin matrices and 
${\bf \tau_{\mu=0}} \equiv {\bf 1} $. 

Then, in the physical subspace defined by Eqs. (\ref{CONST1},\ref{CONST2})
the Hamiltonian (\ref{HM}) takes the form 
\begin{equation}
\tilde{H}= -t\sum_{<ij>,\sigma \sigma_1 \sigma_2}
z_{i,\sigma \sigma_1}^{*}c_{i,\sigma_1}^{\dagger}
c_{j,\sigma_2}z_{j,\sigma_2 \sigma} 
+ U\sum_{i}d_{i}^{2} \label{SB}
\end{equation}
where  
\begin{eqnarray}
{\bf z_{i}} &=& {\bf L_{i}}(e_i{\bf p_i} + {\bf \tilde{p}_i} d_i) {\bf R_i}
\label{zdef} \\
{\bf L_i} &=& \left\lbrack (1-d_i^2){\bf 1} - {\bf p_{i}^{*}p_{i}} 
\right\rbrack^{-1/2} \\
{\bf R_i} &=& \left\lbrack (1-e_i^2){\bf 1} - {\bf \tilde{p}_{i}^{*}
\tilde{p}_{i}} 
\right\rbrack^{-1/2}
\end{eqnarray}
The matrices ${\bf L_i}$ and ${\bf R_i}$ guarantee the correct behavior
in the limit $U \rightarrow 0$ within the mean-field approximation and
${\bf \tilde{p}_i}=\hat{T} {\bf p_i} \hat{T}^{-1}$ 
is the time-reversal transformed of ${\bf p_i}$.

The matrix elements of ${\bf z_i}$ can be calculated by transforming to 
a diagonal representation for the ${\bf p_i}$ (see Appendix).
The resulting effective one-particle Hamiltonian 
describes the dynamics
of particles which upon hopping between sites are subjected to
a modulation of their spin amplitude and spin direction, respectively.
It can be diagonalized by the transformation
\begin{equation}
c_{i,\sigma}=\sum_{k}\Phi_{i,\sigma}(k)a_{k}
\end{equation}
where the orthogonality of the transformation requires 
\begin{equation}\label{CONST3}
\sum_{i,\sigma}\Phi^{\ast}_{i,\sigma}(k)\Phi_{i,\sigma}(q)=\delta_{kq}.
\end{equation}

Given a system with $N_{el}$ particles we finally obtain for the
total energy
\begin{eqnarray}
E_{tot}&=&-t\sum_{<ij>,\sigma \sigma_1 \sigma_2}
z_{i,\sigma \sigma_1}^{*} z_{j,\sigma_2 \sigma}\sum_{k=1}^{N_{el}}
\Phi^{\ast}_{i,\sigma_1}(k)\Phi_{j,\sigma_2}(k) \nonumber \\
&+&U\sum_{i}d_{i}^{2} \label{E1}
\end{eqnarray}
which has to be evaluated within the constraints (\ref{CONST1},\ref{CONST2},
\ref{CONST3}).
This is achieved by adding these constraints quadratically to Eq. (\ref{E1})
following the procedure already applied in the Gutzwiller limit \cite{GSH}.
The resulting energy functional then has to be minimized
with respect to the fermionic and bosonic fields which is most
conveniently done by using a standard conjugate
gradient algorithm since the gradients of the energy functional can be 
calculated analytically. 
In order not to end up in pathological side minimas we have generally 
started the minimization from an HF Ansatz for the amplitudes
$\Phi_{i,\sigma}(k)$.

\section{Results}
Since in a previous publication \cite{GSH} the unrestricted 
slave-boson approximation has been already applied to
the description of collinear spin structures,  we will restrict here 
to textures with two- and three dimensional spin ordering. 
In this section we discuss the spin structure of vortex, skyrmion and 
elliptical domain wall textures which turn out to be stable energy minima 
within our slave-boson approach.
Obviously on finite lattices one has to use open boundary conditions 
in order to describe higher dimensional spin structures and the cluster
sizes we are considering in the following are ranging from $6\times 6$ up to
$10\times 10$.

The incorporation of transversal spin degrees of freedom allows for
the definition of spin currents $\nabla {\bf j} = -\partial_t{\bf S} $.
The flow direction of these currents is along the bonds of the lattice,
however, they are additionally vectorial in spin space and within
the present approach we obtain for the i-th component of the spin current 
flowing between sites $<nm>$:
\begin{equation}
j^{i}_{n m} \sim Im \sum_{\sigma_1 \sigma_2 \sigma_3 \sigma_4}
\sum_{k=1}^{N_{el}}
\Phi^{\ast}_{n,\sigma_1}(k) \tau^i_{\sigma_1 \sigma_2}
z_{n,\sigma_3 \sigma_2}^{\ast}z_{m,\sigma_4 \sigma_3}
\Phi_{m,\sigma_4}(k)
\end{equation}
where $\tau^i$ are the Pauli matrices and the hopping factors
$z_{n,\sigma \sigma'}$ are defined in Eq.\ (\ref{zdef}).
The i-th component of ${\bf j}_{nm}$ can be thought of as measuring
the spin-twist in the orthogonal directions $l,k \ne i$ so that the
total current components  ${j^i_n}$, 
which are plotted in the results,  visualize the direction of 
maximal twist in the spin components $lk \ne i$ at lattice site n. 

\subsection{Vortex States}
The structure of vortex solutions, where the magnetization rotates in a plane
by some multiples of $2\pi$ around the localized holes, 
has already been studied in Ref. \cite{BISHOP}
within unrestricted HF theory. 
We also obtain vortex states as local minima of the energy 
functional (\ref{E1}), where the  total energy
is about $5\%-10\%$ lower than in the HF approach, depending on lattice
size and on-site repulsion U. However, for one hole away from half-filling
vortex solutions are always higher in energy than the conventional 
N\'{e}el ordered spin polaron. Moreover, their total energy increases
logarithmically with the cluster size as a consequence of the twist between
neighboring magnetization vectors in agreement with Ref. \cite{BISHOP}.

\begin{figure}
{\hspace{0.7cm}{\psfig{figure=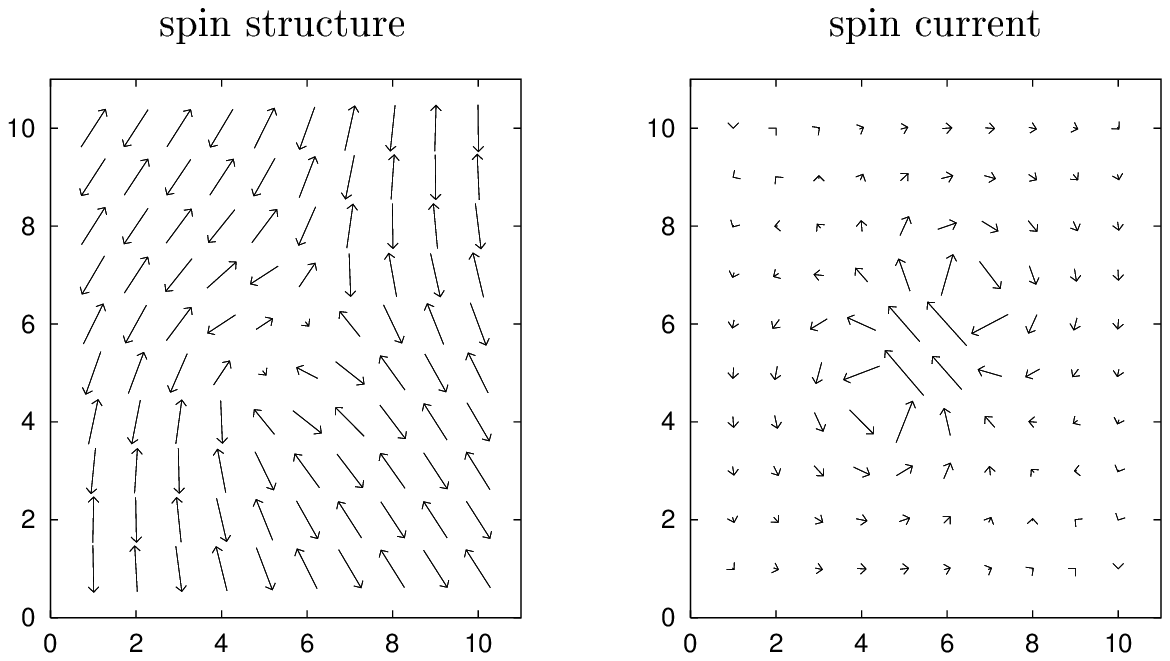,width=6.5cm}}}

\vspace*{4.5cm}

{\small FIG. 1. Spin structure and spin currents for a vortex-antivortex
pair on a $10 \times 10$ lattice. The Hubbard on-site repulsion
is $U=10t$.}
\end{figure}

This logarithmic divergency can be compensated when two holes
form a vortex-antivortex pair. According to our calculations the
vortex cores are located at the center of diagonally next nearest
neighbor plaquettes, thus separated by the plaquette where
the two holes are localized. Fig.\ 1 shows the spin structure and
spin current of such a vortex-antivortex pair on a 
$10 \times 10$ cluster and  $U=10t$. All the spins lie in the
xy-plane which means that only the z-component of the spin current has a
non zero contribution.
The spin field of this texture can be described by
\begin{equation} \label{VAEQ}
{\bf S}_i = S_0 e^{i {\bf Q R}_i} \lbrack \cos(\phi_1-\phi_2){\bf e_x}
+ \sin(\phi_1-\phi_2){\bf e_y}\rbrack
\end{equation}
where $\phi_{1}$, ($\phi_{2}$) refer to the angles between x-axis and 
the vectors connecting  vortex (antivortex) core and site ${\bf R}_i$.
The AF wave vector is denoted by ${\bf Q}$.

Concerning the stability of the vortex-antivortex pair we obtain for
the $10 \times 10$ cluster and $U=10t$ a binding energy of 
$E^{b}=-0.047t$ with respect to the N\'{e}el type bipolaron.  
Taking into account the far field energy, calculated within the
xy-model for the spin structure Eq.\ (\ref{VAEQ}) and an exchange
constant of $J=4t^2/U=0.4t$ still results in a negative binding energy of 
$E^{b}_{tot}=-0.016t$.

We note that the absolute value of $E^{b}_{tot}$ slightly 
decreases for smaller cluster sizes since the boundary spins do not
properly adjust to the solution Eq.\ (\ref{VAEQ}). One can therefore safely
conclude that vortex-antivortex pairs are also stable in the thermodynamic
cluster limit.

\subsection{Skyrmions}
On a discrete 2-dimensional AF lattice the spin structure of skyrmions, 
originally obtained 
as solutions of the O(3) non-linear $\sigma$-model \cite{BELAVIN}, has the
form \cite{SHRAIMAN,GOODING}
\begin{eqnarray}
S_x &=& (-1)^{i_x+i_y}\frac{\lambda i_x}{i_x^2+i_y^2+\lambda^2}\nonumber\\
S_y &=& (-1)^{i_x+i_y}\frac{\lambda i_y}{i_x^2+i_y^2+\lambda^2} \label{SKF}\\
S_z &=& (-1)^{i_x+i_y}\frac{1}{2}\frac{i_x^2+i_y^2-\lambda^2}
         {i_x^2+i_y^2+\lambda^2}\nonumber
\end{eqnarray}
where $\lambda$ denotes the core size of the skyrmion and its center is
located at $i_x=i_y=0$.

In order to enhance convergence we initialized our minimization 
procedure with (non self-consistent) HF 
wave functions  corresponding to the spin fields Eqs.\ (\ref{SKF}).

Despite intensive search we could not obtain skyrmion states for one
hole doped in the half-filled system. These solutions always converged
towards a spin-polaron embedded into a collinear AF N\'{e}el state as already
observed in Ref. \cite{BISHOP}.

\begin{figure}
{\hspace{0.7cm}{\psfig{figure=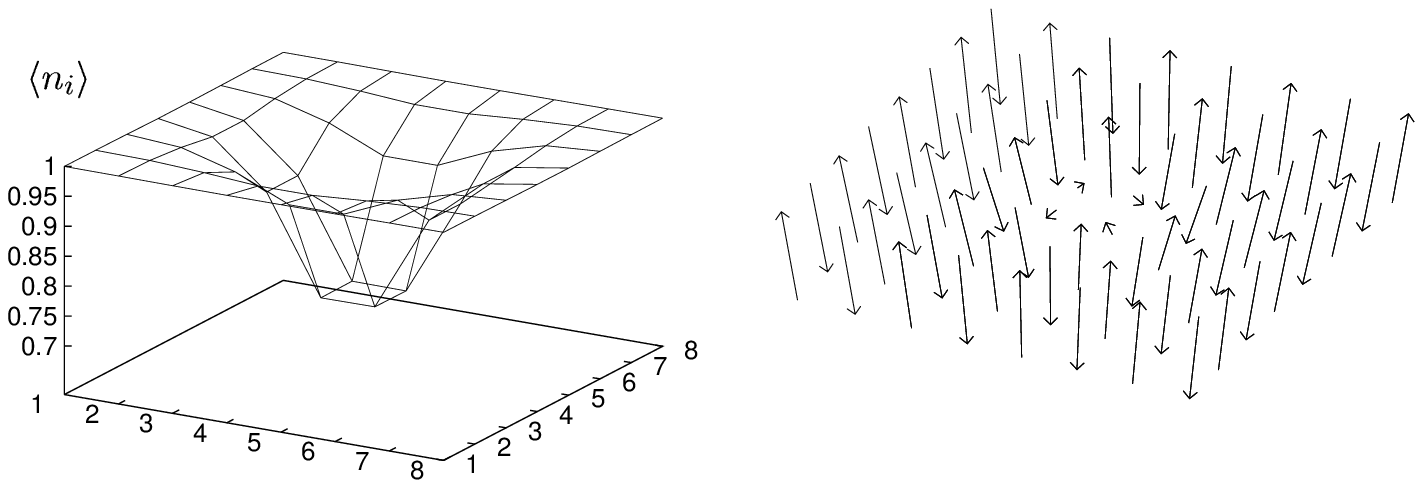,width=5.5cm}}}

\vspace*{4cm}

{\small FIG. 2. Charge- ($\langle n_{i} \rangle$) and Spin-distribution for
         a skyrmion texture on a $8\times 8$ lattice for
         $U=10t$. }
\end{figure}
\end{multicols}

\begin{figure}
{\hspace{3cm}{\psfig{figure=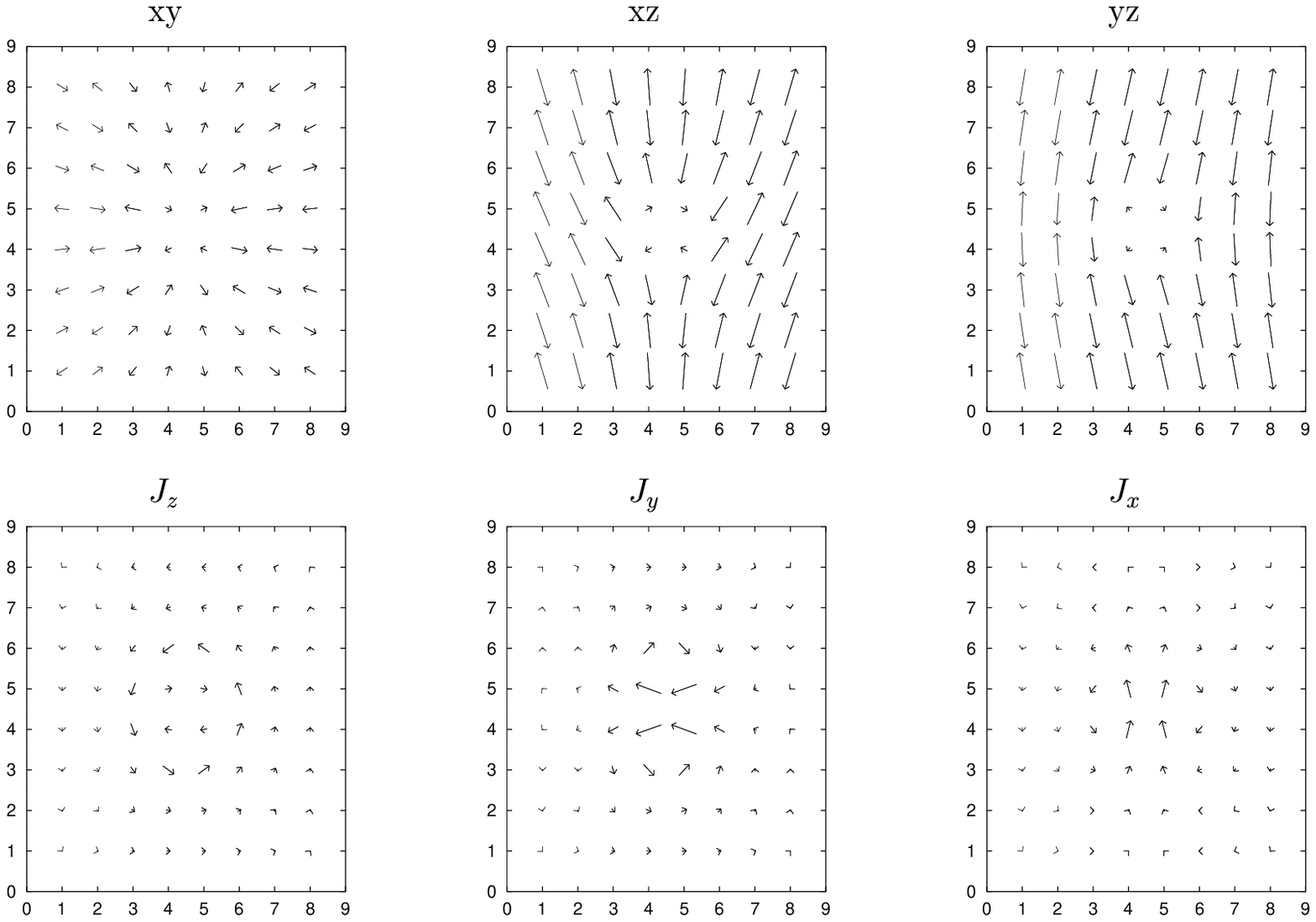,width=10.0cm}}}

\vspace*{4.7cm}

{\small FIG. 3. xy-, yz- and xz-projections of the spins together with the
         respective spin currents for the skyrmion shown in Fig.\ 2 }
\end{figure}
\begin{multicols}{2}
The situation changes when removing two particles from the half-filled
system. Fig.\ 2 displays the charge- and spin structure in case of 
62 particles on a $8\times 8$ lattice for $U=10t$. 
The two holes then are localized on a $2\times 2$
plaquette at the skyrmion center and in their vicinity spins show a
remarkable deviation from the z-direction.
This is more easily seen in Fig.\ 3 where we have plotted the
xy-, xz- and yz- spin projections, respectively,  
together with the corresponding spin currents.
The xy-spin components rotate by $360^{\circ}$ around the skyrmion
center resulting in a circular spin current for $j^z$.
However, all current components strongly decay for sites far away from
the skyrmion center indicating that the core size parameter
$\lambda$ is small. Upon fitting our solutions to the skyrmion field
Eq.\ (\ref{SKF}) we obtain $\lambda \approx 0.7$ for $U=10t$ and cluster sizes
$6 \times 6$, $8 \times 8$ and $10 \times 10$, respectively.
This already indicates that the skyrmion state should survive the
limit of large clusters.

To assess the question of stability in more detail we have also
calculated the total energy using skyrmionic boundary conditions \cite{HAAS}.
These are defined through an exchange field $J {\bf \cal S}(R,\lambda)$
to which the spins at the boundary are coupled and ${\bf \cal S}(R,\lambda)$
has the form of Eq.\ (\ref{SKF}).
For the exchange constant we take the strong-coupling value of
the Hubbard model $J=4t^2/U$. The total energy then 
is evaluated as a function of $\lambda$ and in the result we substract the 
energy contribution of the exchange field. This energy has to be
compared with the corresponding value of a collinear bipolaron where
two holes are localized on neighbored sites within the N\'{e}el
ordered system. The results are plotted in Fig.\ 4 again for 
$U=10t$ and three different cluster sizes.  
As can be seen all curves display a clear minimum at some value of $\lambda$
indicating the presence of a stable skyrmion solution with a significant
lower energy than the collinear bipolaron. It should be mentioned that these
minima for the corresponding one-hole doped systems are always at
$\lambda=0$, i.e. the configuration of a conventional spin-polaron.
From the results shown in Fig.\ 4 one further sees that a lowering
of energy with respect to the AF bipolaron is already obtained for
$\lambda=0$ indicating that despite the imposed N\'{eel} boundaries
the system has a skyrmion like core. This energy shift increases
with the system size since the central spins of large lattices
can more properly adjust to the skyrmion state. From the fact that
the $\lambda=0$ results for the $8\times 8$ and $10 \times 10$ coincide
within the numerical error we conclude that also in the thermodynamic
limit the skyrmion solution should survive.
Also plotted in Fig.\ 4 with filled symbols are the energy differences 
between bipolarons and skyrmions calculated with open boundary conditions.
As already mentioned one obtains the same core parameter $\lambda$
for all three cluster sizes which agrees with the position of the
minimum of the solid line ($10 \times 10$ system). Also this feature
demonstrates that our largest cluster should already correctly describe
the skyrmion structure of infinite clusters. Since the skyrmion on
small lattices is very much influenced by the boundary conditions 
the minimum of the dotted curve ($6 \times 6$ cluster) is shifted to
a higher value of $\lambda \approx 1$. 
 
\begin{figure}
{\hspace{1.5cm}{\psfig{figure=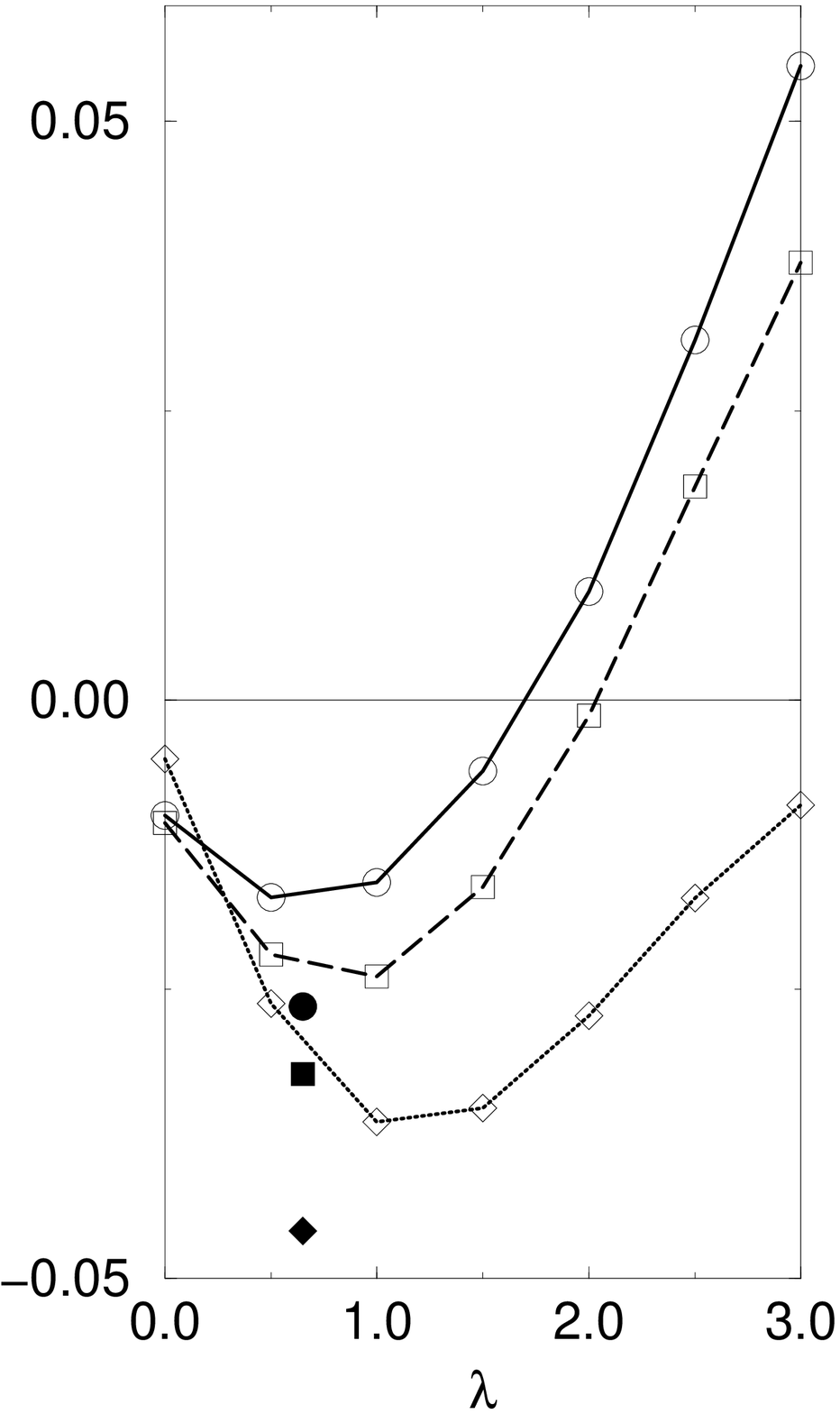,width=4.0cm}}}

{\small FIG. 4. Total energy of a two-hole-skyrmion with respect to the energy
         of a N\'{e}el-type bipolaron as a function of the skyrmion core
         size parameter $\lambda$. The boundary spins have been coupled
         to the skyrmion solution Eq.\ (\ref{SKF}) via a mean-field exchange 
         field. Solid line and circles: $10 \times 10$ lattice;
         Dashed line and squares: $8 \times 8$ lattice; Dotted line
         and diamonds: $6 \times 6$ lattice. The full symbols mark
         the energies for open boundary conditions. $U=10t$.}
\end{figure}

In order to compare the stability of skyrmion states with the 
vortex-antivortex solutions, one has to take into account the
far field contribution to the total energy. 
Considering again a $10 \times 10$ system and $U=10t$ the skyrmion
binding energy with respect to the N\'{e}el-type bipolaron 
(cf. Fig.\ 4) is $E^{b}_{tot}=-0.0265t$.
Including the far field energy (taking again $J=0.4t$) one
obtains $E^{b}_{tot}=-0.0071t$ which is approximately half the value
of the vortex-antivortex binding energy. 

\subsection{Elliptical domain walls}
The possibility that charge-spin coupling in correlated systems 
may induce the formation of elliptical domain walls was proposed by an 
analysis of the Landau free-energy functional for coupled charge- and
spin density waves \cite{ZACHAR}. These are coplanar spin structures 
where the spin components in the first harmonic are modulated as
\begin{eqnarray}
S_x({\bf r})&=& S_0 e^{i {\bf Q r}}\cos(\alpha) \cos({\bf qr}) \nonumber \\
S_y({\bf r})&=& \pm S_0 e^{i {\bf Q r}}\sin(\alpha) \sin({\bf qr}) \label{EDW}
\end{eqnarray}
The eccentricity of the elliptical domain wall is determined by
$\alpha$ and ${\bf Q, q}$ correspond to the wave vectors of commensurate AF 
and the domain wall periodicity, respectively.
The case $\alpha=\pi/4$ describes an ideal spiral solution whereas $\alpha=0$
reduces the spin structure to a collinear 'classical' domain wall.

We have used (non-self consistent) HF states corresponding to Eq.\ (\ref{EDW}) 
as starting fields for our minimization for different values of 
$\alpha$. Only vertical domain walls have been considered and 
periodic boundary conditions where applied in the x- and y-direction, 
respectively. In case of a completely filled domain wall 
(i.e. one hole per site along the wall) we only found the
collinear solutions whereas coplanar structures become stable
for half-filled walls.

\begin{figure}
{\hspace{-0.2cm}{\psfig{figure=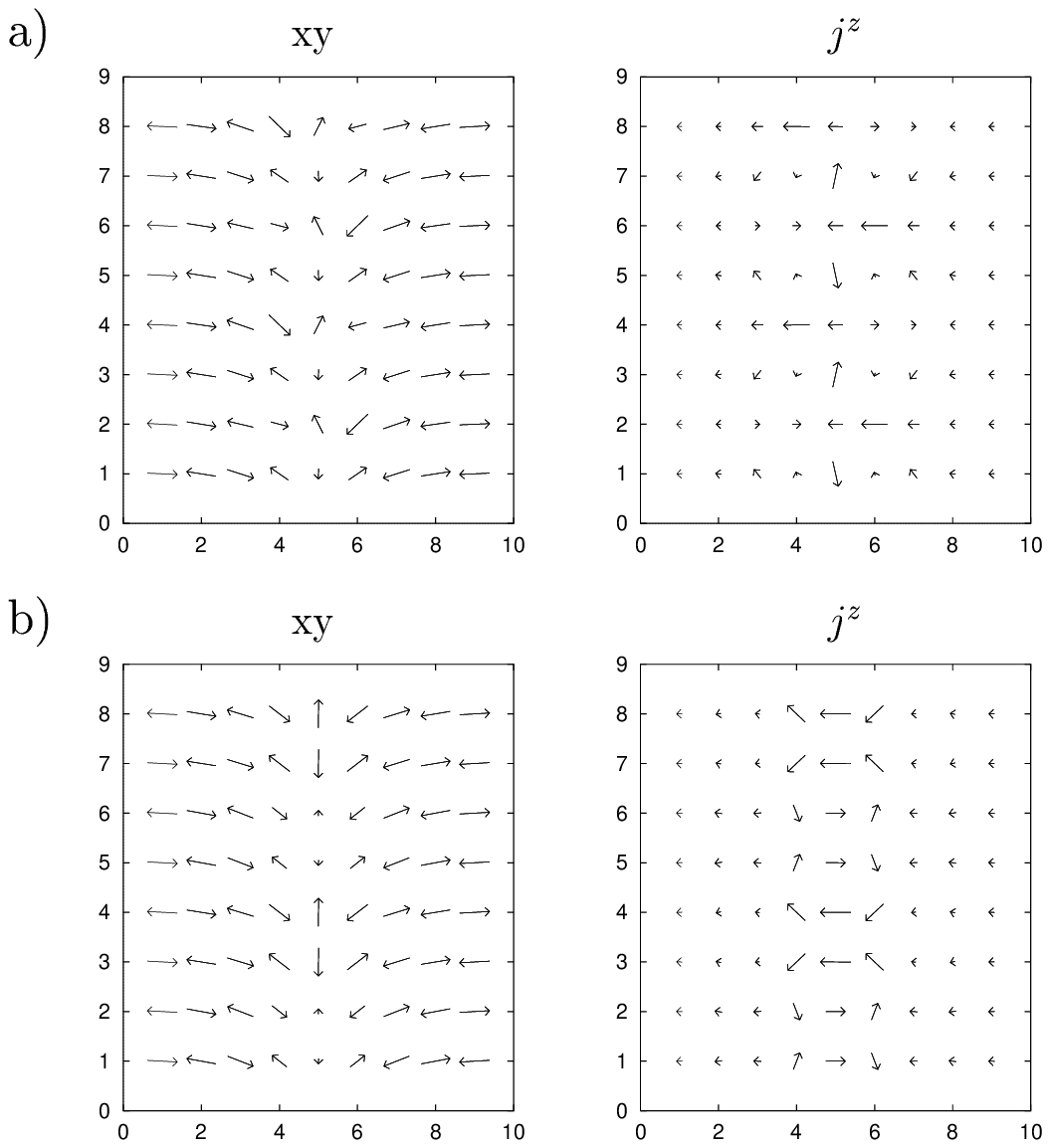,width=7cm}}}

\vspace*{4.7cm}

{\small FIG. 5. Two possible spin structures for elliptical domain walls 
         together
         with the corresponding spin currents. The on-site repulsion is
         $U=6t$ and periodic boundary conditions in x- and y-direction
         have been used.
         Shown are the results for a $9 \times 8$ lattice doped with
         4 holes. }
\end{figure}

This is shown in Fig.\ 5 for a $9\times 8$ lattice with 68 particles
where we plot two kinds of possible spin structures which are 
local minima of the energy functional Eq.\ (\ref{E1}). Also shown
are the respective spin currents (only z-components since the magnetization
vector is completely in the xy-plane).
Spin fields and currents display a quadrupled structure along the
wall which is a necessary condition for stability within any mean-field
approach \cite{OLES}.
Due to the stripe charge structure the current flows are more complex
than for the elliptical solution Eq.\ (\ref{EDW}) which predicts currents
flowing in a single direction orthogonal to the wall.
Instead we observe also currents along the wall (Fig.\ 5a) and a
vortex-antivortex structure in Fig.\ 5b.

Although the elliptical stripes are local minima of the energy functional
Eq.\ (\ref{E1}) they are slightly higher in energy ($\approx 5\%$) than
collinear domain walls. These have been shown to correspond to the
ground state when one takes long-range Coulomb interactions into
account \cite{SCCG}. However, we do not expect significant differences
of a long-range contribution to collinear and elliptical stripe
solutions.

The structures shown in Fig.\ 5 correspond to systems with
hole doping $1/18$ and we could not obtain elliptical solutions
for higher doping. However, one can also study the  very low doping limit
upon using open boundaries in the x-direction.
Indeed in this case elliptical half-filled stripes become favored with
repect to collinear domain walls for not too large values of the on-site
repulsion U ($<9t$).

\begin{figure}
{\hspace{-0.2cm}{\psfig{figure=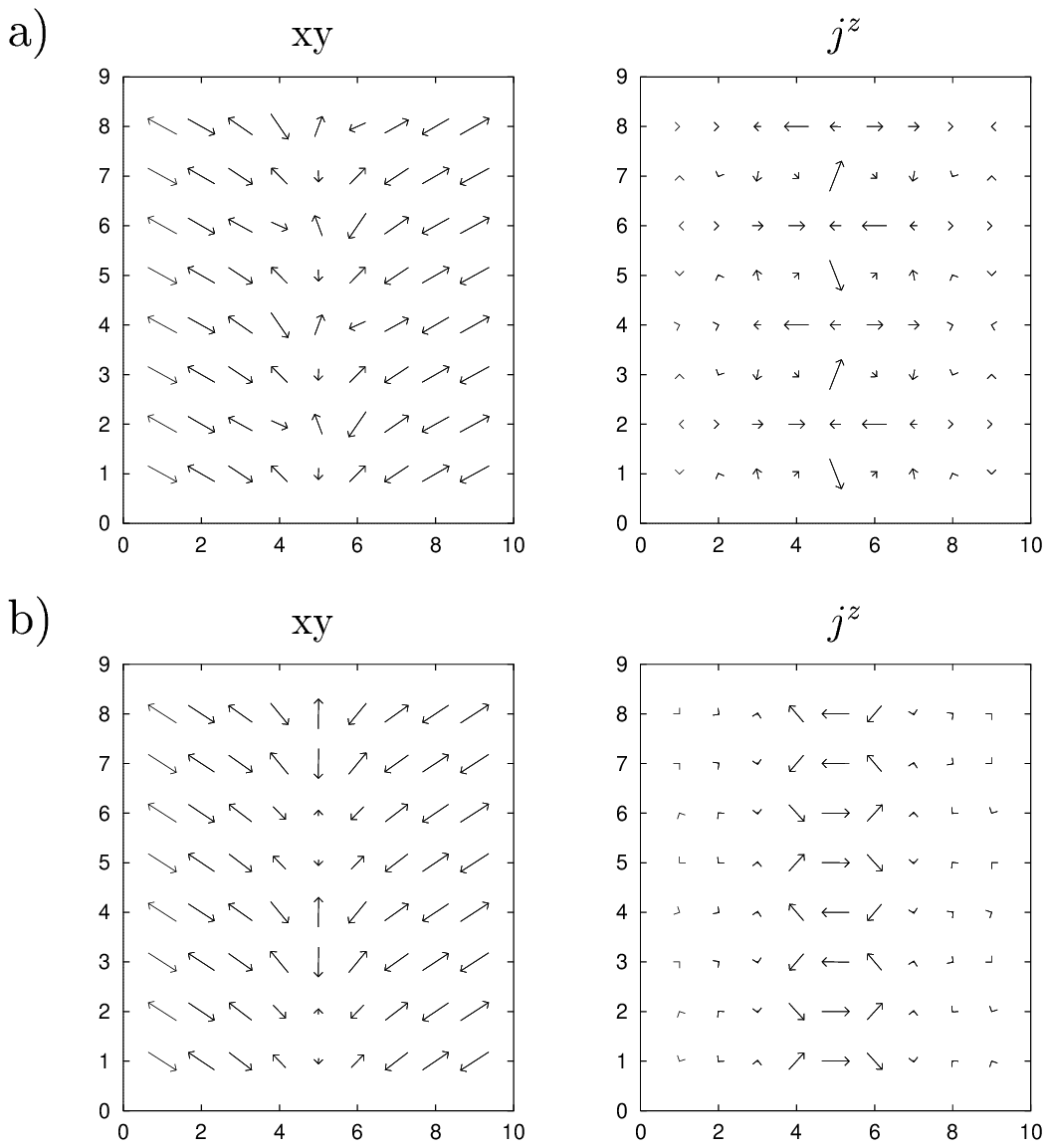,width=7.0cm}}}

\vspace*{4.7cm}

{\small FIG. 6. The same as in Fig. 5 but now for open boundaries in 
                x-direction. }
\end{figure}

In Fig.\ 6 we show the same spin structures as in Fig.\ 5 but for open
boundaries in x-direction.  
The main difference between these two figures
concerns the angle of spin rotation $\Theta$ across the 
domain wall which is only $\approx 3/4$ 
of the expected value of $\pi$ according to Eq.\ (\ref{EDW}).
This is due to the fact that the system now can acquire a state
which has zero total spin current, whereas for periodic boundaries
the spiral component always requires a net flow in x-direction.
 
However, since for a regular array of stripes the charge periodicity
and the spin modulation have to be related by $k^{charge}=2 k^{spin}$
\cite{ZACHAR} the domain wall induced kink type spin rotation 
has to be supplemented by an additional spiral field.
Assuming an exponential relaxation of the spin rotation from $\Theta$
to $\pi$ by this spiral field gives an additional energy per site
of $\Delta E \approx \rho_s (\pi - \Theta)^2 \lambda /L$ where 
$\rho_s=J S^2$ is the spin stiffness, $\lambda$ defines the length
scale of relaxation and L is the stripe separation. 
Consequently this additional energy can become small enough in
the low doping regime so that elliptical stripe configurations are more
stable than collinear solutions for small hole concentrations.

\section{Conclusion}
Summarizing, we have presented the structure of spin textures 
which we obtained by applying an unrestricted slave-boson 
mean-field approximation in its spin-rotation invariant form to the
2D Hubbard model.
This approach is suited for calculating the charge and spin
distribution of electrons (holes) in  inhomogeneous and
strongly correlated systems since it incorporates correlation effects
beyond the standard unrestricted HF theory.
These correlation effects turn out to be especially important for the
interaction among holes in the 2-D Hubbard model where
the HF approximation strongly underestimates their attractive potential
\cite{GSH}.

Including the effect of transversal degrees of freedom we have shown 
that within our slave-boson mean-field approach two 
holes in the half-filled two-dimensional Hubbard model are bound by
forming a vortex-antivortex pair, oriented along the diagonal
direction. This texture has significantly
lower energy than a conventional collinear bipolaron, also when
the far field contribution is taken into account.
Additionally we have found that skyrmion states can be
stable even on discrete lattices. We attribute this to the
inclusion of correlation effects within our approach, since
unrestricted Hartree-Fock theory cannot account for skyrmions
as self-consistent solutions (or local minima of energy).
Indeed it has been also observed by the authors of
Ref. \cite{HAAS} that within the tJ-model a semiclassical
description cannot account for the occurence of skyrmions but
that it is the quantum fluctuations which stabilize this texture.
 Considering the formation of elliptical domain walls it turned out that
these structures only appear for half-filled walls in the low doping
regime. Similar to the collinear stripes they are stabilized by
a quadrupling of the period along the wall.
This can be realized either by alternating on-wall spin currents
(Fig.\ 5a) or by forming a vortex-antivortex structure (Fig.\ 5b).

\acknowledgements
We would like to thank V. Hizhnyakov for valuable discussions
and a critical reading of the manuscript.

\appendix
\section*{}
Our purpose is to calculate the matrix elements of the hopping matrices 
${\bf z}_i$ by transforming to a diagonal
representation of the matrix ${\bf p_i}$ (Eq.\ (\ref{PI})) according to
\begin{equation}
T({\bf p_i})\hat{=}{\bf \chi_i}^{-1}{\bf p_i} {\bf \chi_i}
= \left( \begin{array}{cc} \lambda_i^{+} & 0 \\
0 & \lambda_i^{-} \end{array}\right)
\end{equation}
where the eigenvalues are given by $\lambda_i^{\pm}=\frac{1}{2}(p_{i,\uparrow}
+p_{i,\downarrow})\pm \frac{1}{2}\sqrt{(p_{i,\uparrow}-p_{i,\downarrow})^2
+2(p_{i,x}^2+p_{i,y}^2)}$.
The transformation matrix ${\bf \chi}_i$ reads as
\begin{eqnarray}
{\bf \chi}_{i} &=& \left( \begin{array}{cc} 
\alpha_{i}^{+} & \alpha_i^{-}\\
\beta_{i}^{+} & \beta_i^{-} \end{array}\right) \\
\alpha_i^{\pm}&=&-\frac{1}{\sqrt{2}}\frac{p_{i,x}-i p_{i,y}}
{\sqrt{(p_{i,\uparrow}-\lambda_i^{\pm})^2+p_i^2/2}}\\
\beta_i^{\pm}&=& \frac{p_{i,\uparrow}-\lambda_i^{\pm}}
{\sqrt{(p_{i,\uparrow}-\lambda_i^{\pm})^2+p_i^2/2}}
\end{eqnarray}
Applied to the hopping matrix this transformation yields
\begin{eqnarray}
{\bf z}_i &=&{\bf \chi_i}T({\bf z_i}) {\bf \chi_i} \nonumber \\
T({\bf z_i}) &=& T({\bf L_{i}})\lbrack e_i T({\bf p_i}) + 
T({\bf \tilde{p}_i}) d_i\rbrack T({\bf R_i})
\label{ZTRAF}
\end{eqnarray}
and since $T({\bf p_i})$ is diagonal it is straightforward to evaluate
$T({\bf L_{i}}), T({\bf R_i})$ and $T({\bf \tilde{p}_i})$ as
\begin{eqnarray}
T({\bf L_{i}}) &=& \left( \begin{array}{cc} 
\frac{1}{\sqrt{1-d_i^2-(\lambda^{+})^2}} & 0\\
0 & \frac{1}{\sqrt{1-d_i^2-(\lambda^{-})^2}} \end{array}\right) \\
T({\bf R_{i}}) &=& \left( \begin{array}{cc} 
\frac{1}{\sqrt{1-e_i^2-(\lambda^{-})^2}} & 0\\
0 & \frac{1}{\sqrt{1-e_i^2-(\lambda^{+})^2}} \end{array}\right) \\
T({\bf \tilde{p_{i}}}) &=& \left( \begin{array}{cc} 
\lambda^{-} & 0\\
0 & \lambda^{+} \end{array}\right) 
\end{eqnarray}

\end{multicols}

\end{document}